\documentclass[prb,aps,preprint]{revtex4}
\usepackage{graphicx,amsmath}

\begin{document}

\title{Dithiocarbamate Anchoring in Molecular Wire Junctions: A First Principles Study}

\author{Zhenyu Li}
\author{Daniel S. Kosov}

\address{Department of Chemistry and Biochemistry,
 University of Maryland, College Park, 20742}


\begin{abstract}
Recent experimental realization [J. Am. Chem. Soc., 127 (2005) 7328] of
various dithiocarbamate self assembly on gold surface opens the possibility for 
use of dithiocarbamate linkers to anchor molecular wires to gold electrodes.
 In this paper, we explore this hypothesis computationally. We computed the
 electron transport properties of 
 4,4$^\prime$-bipyridine (BP), 
 4,4$^\prime$-bipyridinium-1,1$^\prime$-bis(carbodithioate) (BPBC),
 4-(4$^\prime$-pyridyl)-peridium-1-carbodithioate (BPC)
 molecule junctions based on the  density functional theory and non-equilibrium 
Green's functions.  We demonstrated that the stronger molecule-electrode coupling 
associated with the  conjugated dithiocarbamate linker broadens transmission resonances near the Fermi energy. 
The broadening effect along with the extension of
  the $\pi$ conjugation from the molecule to the gold electrodes lead to enhanced electrical conductance for  BPBC molecule.
  The conductance enhancement factor is as large as 25 at applied voltage bias 1.0 V.
  Rectification behavior is predicted for BPC molecular wire junction, which has the asymmetric anchoring groups.
\end{abstract}

\maketitle

\section{Introduction}
Considerable experimental and computational efforts have been devoted in recent years
to the problem of molecular electronics. \cite{Nitzan03,Joachim05}
These efforts have been spurred on by the anticipation for
breakthrough technological applications.
When an electronic devices shrink to molecular level,
the molecule-electrode contact becomes a major
part of the device and  chemical details of the bonding 
becomes pivotal 
for the device electron transport properties.\cite{Xue03, ke0497, Basch05, Ke05} 
To date, the most widely used molecular wire
junctions have been  formed by thiolated molecules assembled between 
gold electrodes.
There is growing experimental awareness of the fact that one of the 
most important challenges in molecular-scale electronics is 
the ill-defined bonding between molecular wires and gold electrodes.
The aim of  the reproducible experimental measurements has
become the search for better molecule-electrode bonding.\cite{Siaj0588, Tulevski0591, Tivanski0598, Guo0656, Venkataraman0600,He06} 
Most of recent experiments has attempted to circumvent 
shortcomings of thiol linkage  by using carbenes on transition metals \cite{Tulevski0591}
and on metal carbides\cite{Siaj0588}, by inserting molecular wires into nanogaps in single-walled carbon nanotubes \cite{Guo0656}, or by fabricating metal-free silicon-molecule-nanotube molecular devices.\cite{He06}
But being an excellent conductor with mature manipulation technics,  gold remains one of the
most attractive electrode material for molecular electronics. 

It has been recently shown  by Zhao \textit{et al.}\cite{Zhao05}
that dithiocarbamate formation provides a reliable technology for
conjugating secondary amines onto metal surfaces to form strongly
absorbed  molecular ligands which are stable under various types
of the environmental stress. This result suggests a
possibility of connection of molecular wires to gold electrodes
via dithiocarbamate group instead of thiol group. In fact, the distance between the
sulphur atoms in the dithiocarbamate group is  nearly ideal for
epitaxial adsorption onto Au surface, and a variety of secondary
amines can condense with dithiocarbamate linker onto Au surfaces
at room temperature,\cite{Zhao05} which makes the dithiocarbamate
anchoring structure to be easily fabricated and 
characterized. On the other hand, the dithiocarbamate anchoring
group provides strong molecule-electrode coupling, and thus the
possibility of more stable junctions with enhanced conductance.

Theoretically, stronger molecule-electrode coupling leads to more
significant manifestation of the effects associated with the
molecule-electrode interface. The substantial charge transfer
may occur between the molecule and the metal electrodes even in the
absence of the applied field upon initial chemisorption and it
will determine the transport  properties of the molecular wire
junctions. The stronger is the molecule-electrode coupling, the
larger is  the mixing between the discrete molecular levels and
the continuum of the metal electronic states and the larger is the
broadening of the resonances in the electron transmission
probability. The dithiocarbamate anchoring group thus provides an
ideal testing ground for computational and experimental
explorations of the role of such pivotal effects in the transport
properties of molecular wires.

Dithiocarbamate assembly on the gold surface also opens
interesting possibility for creating rectifying molecular wire
junctions where electrons flow along one preferential
direction.\cite{Metzger03} A number of donor-acceptor type
molecular rectifiers has been proposed, \cite{Aviram7477,
Metzger03, Morales05, Elbing05, Derosa03} in which 
the electron density asymmetry (like
in semiconductor p-n junctions) causes the current to flow in one
direction. Rectification effect could also be achieved by
adjusting the molecule-electrode bond length as it was suggested
theoretically,\cite{Taylor02} but it is difficult to realize such
kind of the rectifying molecular wire junctions in a controllable
way. We propose that molecular rectification can be realized in a  more
practical way by choosing different anchoring group for the left 
and the right sides. 

The aim of this paper is to explore computationally the role of
dithiocarbamate anchoring group on transport properties of
molecular wire junctions. In this paper, we consider three systems
as  prototype molecular wire junctions: two symmetric molecular
wires 4,4$^\prime$-bipyridine (BP) and
4,4$^\prime$-bipyridinium-1,1$^\prime$-bis(carbodithioate) (BPBC),
and one asymmetric junction
4-(4$^\prime$-pyridyl)-peridium-1-carbodithioate (BPC). We choose
the bypyridine based molecular wire junctions because its transport properties
 have been thoroughly  characterized, both experimentally
\cite{Xu03a, Xu03} and theoretically \cite{Tada04,Wu05,Perez05}.
With two nitrogen atoms at the two opposite ends, BP 
is able to connect to gold electrode directly. By repeatedly
forming thousands of molecular junctions between gold scanning
tunneling microscope tip and a gold substrate in solution, Xu
\textit{et al.} \cite{Xu03} found the conductance of BP molecule
near zero bias was about 0.01 $G_0$, where $G_0=e^2/\pi
\hbar=77.48 \mu S$ is the quantum of conductance. Theoretically,
the conductance of directly linked BP molecular wire has been
calculated non-self-consistently by both cluster \cite{Tada04} and
bulk electrode model.\cite{Perez05} Very recently, a more precise
fully self-consistent non-equilibrium Green's functions (NEGF)
calculation has also been performed on this system.\cite{Wu05} In
this paper, we perform combined density functional theory (DFT)
and NEGF calculations on the electron transport properties of
BP,BPBC, and BPC. The electric current at
finite bias voltage is determined self-consistently in our calculations. We find that
dithiocarbamate linkers yield significant broadening of peaks in
the transmission spectra and  improve significantly electron
transport properties of molecular wire junctions.  We suggest and
confirm computationally that  BPC molecular wire junction
rectifies (with factor close to 2) the electric current which
flows through.

\section{Computational Methods}

The electronic transport properties are calculated with the
NEGF technique using TranSIESTA-C
package (version 1.3.0.4). \cite{Brandbyge02} In this  package, the molecular wire junction
 is divided into three regions, left
electrode (L), contact region (C), and right electrode (R). The
contact region typically includes  parts of  the physical electrodes where
the screening  effects take place, to ensure that the charge
distributions in  L and R region correspond to the bulk phases
of the same material. The semi-infinite electrodes
(Au(111)-(3$\times$3) surfaces in our case) are calculated
separately to obtain the bulk self-energy. 

The main loop for  self-consistent NEGF/DFT calculations is described below.
For detailed description and for the technical
details suppressed here we refer to the paper of  Brandbyde
{\it et al.}.\cite{Brandbyge02}
The Green's function is computed by inverting the finite matrix
\begin{eqnarray}
\mathbf{G}(E,V) =
\left[ E \mathbf{S} -
 \left(
\begin{array}{ccc}
H_L +\Sigma_L &  V_L & 0 \\
V_L & H_M & V_R \\
0 & V_R & H_R +\Sigma_R
\end{array}
\right)
\right]^{-1},
\end{eqnarray}
with $V$ being the applied voltage bias.
The matrix product of the Green's function and the imaginary part of the left/right
electrodes self-energy yields the spectral densities $\mathbf{\rho}^{L/R}(E)$.
The spectral densities of the left and right electrodes are combined
together to compute the non-equilibrium, voltage-dependent
density matrix
\begin{equation}
\mathbf{D} = \int dE \left[
\mathbf{\rho}^L(E) f(E-\mu_L)
+ \mathbf{\rho}^R(E) f(E - \mu_R)
\right],
\end{equation}
where $f$ is Fermi-Dirac occupation numbers.
The density matrix is converted into non-equilibrium electron density
\begin{equation}
n(\mathbf{r}) = \sum_{\mu \nu} \phi_{\mu} (\mathbf{r}) D_{\mu \nu}
\phi_{\nu}(\mathbf{r}),
\end{equation}
where $\phi_{\nu}(\mathbf{r})$ atomically localized basis functions.
The nonequilibrium electron density enables us to compute matrix elements
of the Green's function. 
The Hartree potential  is determined through the solution of the Poisson equation with appropriate voltage dependent boundary conditions.
Then this loop of calculations is repeated untill
self-consistency is achieved. 
Once the self-consistent convergence is achieved, 
the transmission
spectrum, which gives the probability for electron with incident energy $E$ to be transfered from  the left electrode to the right,
 is  calculated by 
\begin{equation}
\label{eq:trans}
T(E,V)=Tr[\mathbf{\Gamma}_L(E,V)\mathbf{G}(E,V)
\mathbf{\Gamma}_R(E,V)\mathbf{G}^\dagger(E,V)],
\end{equation}
where
$\mathbf{\Gamma}_{L/R}$ is the coupling matrix.
The integration of  the transmission spectrum yields the electric current:
\begin{equation}
\label{eq:iv} I(V)=\int_{\mu_L}^{\mu_R} T(E,V) \left(f(E-\mu_L) -f(E-\mu_R) \right) dE,
\end{equation}
where 
 $\mu_L =-V/2$ ($\mu_R=V/2$) is the chemical potential of the left(right) electrode.

If we project
the self-consistent hamiltonian onto the Hilbert space spanned by
the basis functions of the central molecule, we obtain the molecular projected
self-consistent hamiltonian (MPSH). The eigenstates of MPSH 
can be associated with poles of the Green's function and thus roughly correspond to the positions of the peaks in the transmission spectrum (\ref{eq:trans}).

The electronic structure is described within the implementation
of DFT in SIESTA computer program, \cite{Soler0245} which
solves the Kohn-Sham equations with numerical atomic basis sets.
Double zeta with polarization (DZP) basis set is chosen for all
atoms except Au, for which single zeta with polarization (SZP) is
used. Our test calculation indicates that using SZP basis set for
Au does not effect the accuracy of our calculations. Core
electrons are modeled with Troullier-Martins nonlocal
pseudopotentials.\cite{Troullier91} Perdew-Zunger local density
approximation (LDA) is used to describe the exchange-correlation
potential.\cite{Perdew8148} By comparing the results from two recent 
calculations with hybrid functional and LDA 
respectively, \cite{Perez05, Wu05} we conclude that 
LDA is adequate in our case. 

\section{Results and Discussion}

\subsection{Structural models of molecular wire junctions}

\begin{figure}
\includegraphics[keepaspectratio,totalheight=12cm]{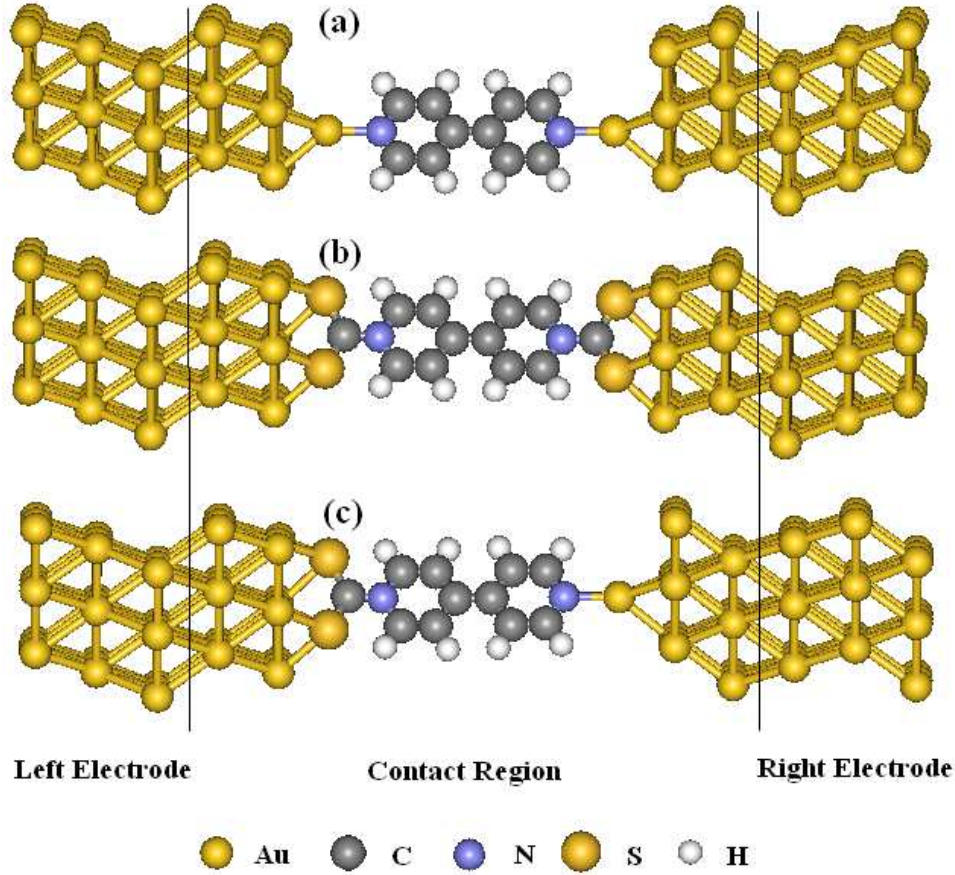}
\caption{The relaxed geometry of Au-BP-Au junctions with different
anchoring groups. a) BP,
b) BPBC,
c) BPC.
}
\label{fig:geo}
\end{figure}

Transport systems considered in this paper are formed by  BP, BPC
or BPBC molecules sandwiched between two Au(111)-(3$\times$3)
surfaces. Fig. \ref{fig:geo} shows the structural models for these
molecular wire junctions. Only one unit cell for the semi-infinite
left or right electrode is plotted, which contains three
Au layers. In the contact region, two Au layers at both left and
right side are included, of which the most left and right layers
are constrained to their theoretical bulk geometry to match the
structure of the  Au(111) surface. 
We assume that all three junctions have coplanar geometry.  This narrows
 down the problem of difference between BP,BPBC, and BPC transport properties
to the role of the dithiocarbamate linkers.
The rest of the contact region
is fully optimized. We also optimized the length of the junctions
by computing the potential energy surface (PES, i.e. the energy of
the system as a function of the distance between the left and the
right electrodes). Every point on the energy surface is calculated
by performing a geometry optimization with constrained
electrode-electrode distance. Then the PES curve is fitted to
determine the correct electrode-electrode distance. 
We also performed test calculations for non-coplanar BP and BPBC.   
Twisting of pyridine rings by $10^\circ$, which corresponds to equilibrium geometry for the
bipyridine junction,\cite{Wu05}  has negligible influence on the transmission spectra. 

Fig. \ref{fig:geo}a shows the optimized geometry of the BP
molecular wire, which is directly connected to two gold
electrodes. When BP molecules are adsorbed on gold surface they
adopt either atop, hollow, or bridge adsorption sites. Depending upon experimental realization BP molecule
may also be connected to the gold surface via apex Au atom. Wu
\textit{et al.} \cite{Wu05} have studied effects of different
interface geometry on the transport properties of BP molecular
wire junction, but focused on the hollow site adsorption. Here we
study the interface geometry with apex Au atom, which is more
likely realized  on the typical experiments where
gold-molecule-gold junctions were formed by repeated pulling of
STM tip from Au surface. \cite{Xu03} This geometrical
configuration also gives the largest value of the conductance.\cite{Ke05, Wu05} 
Therefore our calculations predict the lower limit for the
enhancement of the conductance by means of  dithiocarbamate
anchoring. The structural  model for BPBC molecular wire junction
is shown on Fig.\ref{fig:geo}(b); here each of the four S is
connected to the hollow sites of Au(111)-(3$\times$3) electrode.
This structure resembles the experimental scheme \textbf{7} of
dithiocarbamate ligands formations on Au surface in Fig. 1 of the
paper by Zhao \textit{et al.}.\cite{Zhao05} Fig.\ref{fig:geo}(c)
shows asymmetric BPC molecule sandwiched between two Au
electrodes, in which the left side of the molecule wire is
connected via dithiocarbamate linker and the right side is
connected via N-Au bond to the apex Au atom.

\subsection{Effect of dithiocarbamate anchoring}

\begin{figure}
\includegraphics[keepaspectratio,totalheight=8cm]{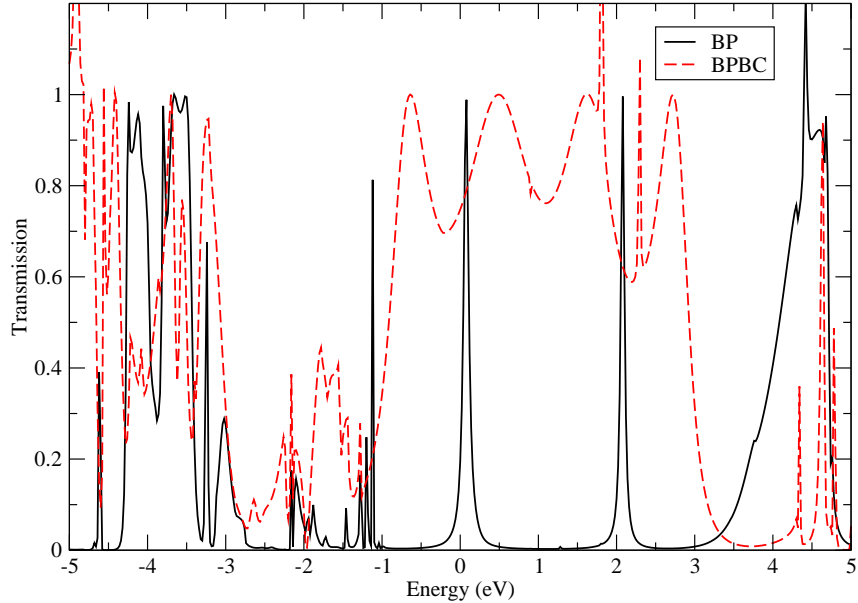}
\caption{The transmission spectrum of molecular wire  junctions. The
average chemical potential of the left and right electrodes is set
to zero.}
\label{fig:trans012}
\end{figure}

To study the effect of dithiocarbamate anchoring, we begin with
calculations of  the transport properties of  BP molecular wire
junction. First, we compute the transmission coefficient $T(E)$ (\ref{eq:trans}). The
zero bias transmission spectrum of BP junction is shown on
Fig.\ref{fig:trans012}. The distinctive feature of this
transmission is the existence of the narrow resonance which
appears very close to  the Fermi level (0.09 eV above $E_F$). This
resonance determines the transport properties of BP molecular
wire  junction under small bias voltages. Two other narrow peaks
are at 2.08 eV above and 1.11 eV below the electrode Fermi energy.
The transmission coefficient at the  Fermi level (i.e. the
conductance at small voltage) is 0.195 $G_0$. This computed value
of the conductance is by order of magnitude larger than the
experimental value of  0.01 $G_0$.\cite{Xu03} But, we notice that the
transmission curve from our calculations is consistent with the
result of the recent first principles calculations on BP
junction,\cite{Wu05,Ke05,Perez05} if the same interface geometry
is considered.  In fact, experimental conductance value can be
obtained theoretically by choosing a suitable model for the
interface geometry.\cite{Wu05} Transmission eigenchannel
decomposition indicates that the zero voltage conductance is
almost entirely determined by a single channel.

\begin{figure}
\includegraphics[keepaspectratio,totalheight=12cm]{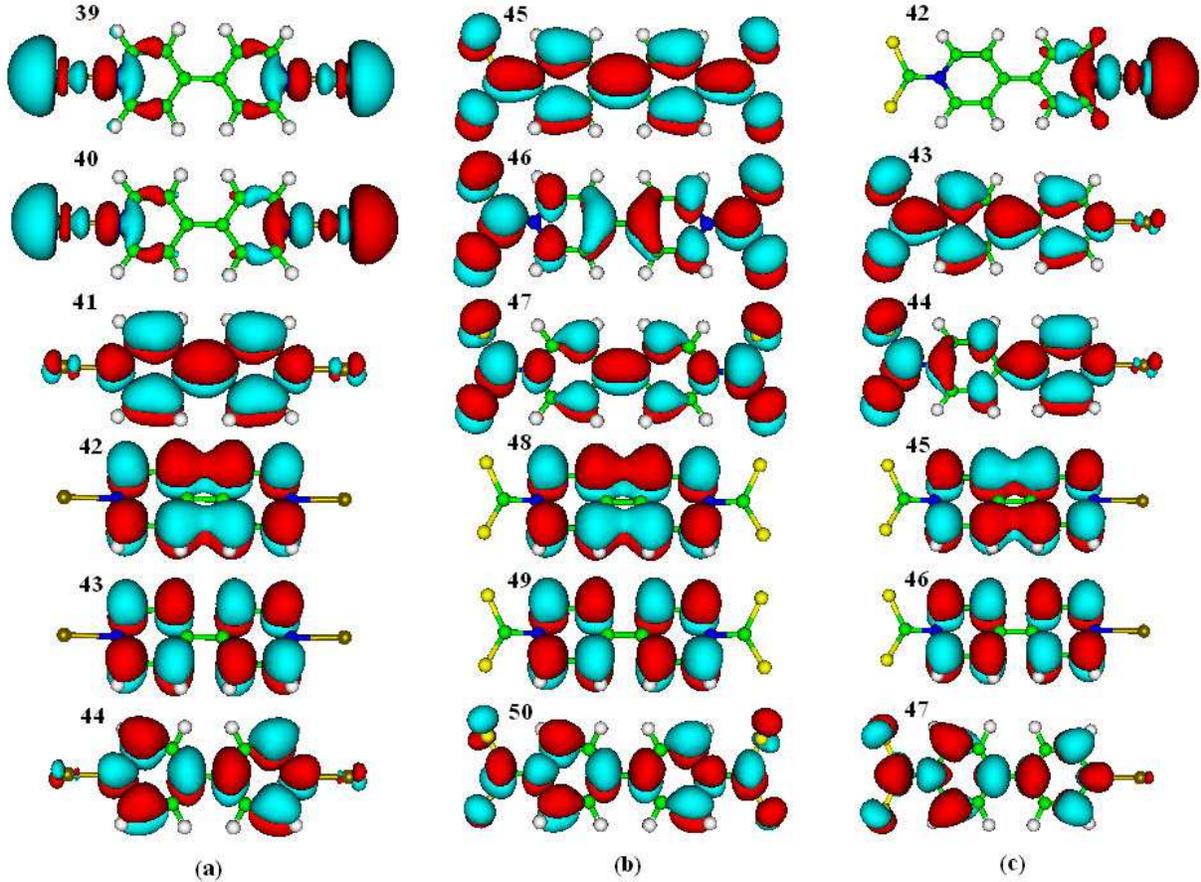}
\caption{Eigenstates of MPSH for (a) BP, (b) BPBC, and (c)
BPC. Anchoring groups and apex Au atom are included for molecular
projection. Refer Fig. \ref{fig:geo} for atom types,  the atoms
are plotted smaller in this figure to present MPSH orbitals better. } \label{fig:MPSH}
\end{figure}

The  transmission spectrum can be further interpreted in terms of MPSH.
The MPSH
orbitals near the average Fermi level are plotted on Fig.
\ref{fig:MPSH}. MPSH orbitals of  BP molecular wire junction
are very similar to the Kohn-Sham orbitals of an isolated BP
molecule except for some additional states which mainly come from
apex Au atoms. It indicates  that the coupling between BP
molecule and apex Au atom is weak, which explains why the
resonances in the transmission spectrum are very narrow. Since the
resonant peaks in the transmission are narrow and well separated, it
is easy to put them into correspondence with the MPSH orbitals by
directly comparing the positions of the resonances with the MPSH
eigenenergies. The MPSH eigenenergies are given in Table
\ref{tbl:energy}. The previously mentioned three peaks correspond
to  the four MPSH orbitals: they are orbitals 39, 40, 41, and 44. Orbital
39 and 40 are almost degenerated. We can see from Fig.
\ref{fig:MPSH}a that all these four orbitals have notable density
distribution on N atoms.

Since the transport properties of BP molecular wire junction
are determined by the narrow resonances in the transmission, 
experimental measurements could be significantly affected by 
stochastic switching and could be very sensitive to variations
of the electrode-molecule interface. The stochastic switching
arises from the structural evolution of the interface
geometry. This is undesirable phenomenon, which leads to the
changes in the conductance by factors of 2-10 depending upon the
scale of the fluctuations. \cite{Ramachandran03, Hu05, Basch05,
Donhauser01} The "robust" transmission spectrum should possess the
following features: a) very broad resonances to reduce the
influence of the current-induced fluctuations of the interface
geometry; b) large transmission probability in the vicinity of the
electrode Fermi energy to make a good conductor from a molecular
wire junction.

\begin{table}
\caption{Eigenvalues of the molecular projected self-consistent Hamiltonian (MPSH) orbitals.
The energy of the average electrode Fermi energy is set to zero. }
\label{tbl:energy}
\begin{tabular}{cccccccc}
\hline\hline
\multicolumn{2}{c}{BP} &&  \multicolumn{2}{c}{BPBC} && \multicolumn{2}{c}{BPC}  \\
\cline{1-2}\cline{4-5} \cline{7-8}
$n$ & Energy(eV) && $n$ & Energy(eV) &&  $n$ & Energy(eV)  \\
\hline
35 & -2.421 && 41 & -2.759 && 38 & -2.408  \\
36 & -2.414 && 42 & -2.746 && 39 & -2.172 \\
37 & -2.354 && 43 & -2.542 && 40 & -2.170  \\
38 & -2.346 && 44 & -2.510 && 41 & -2.121 \\
39 & -1.182 && 45 & -0.851 && 42 & -0.999 \\
40 & -1.134 && 46 &  0.246 && 43 & -0.340 \\
41 &  0.044 && 47 &  1.474 && 44 &  1.234 \\
42 &  1.285 && 48 &  1.786 && 45 &  1.890 \\
43 &  1.806 && 49 &  2.276 && 46 &  2.377 \\
44 &  2.050 && 50 &  2.643 && 47 &  2.677  \\
\hline\hline
\end{tabular}
\end{table}

Dithiocarbamate linker provides strong coupling between the molecular
wire and the gold electrode and it continues $\pi$-conjugation from the
molecule into the gold electrode. Therefore it is very interesting
to see if this anchoring group leads to the "robust" molecular device. From scheme \textbf{7} of Fig. 1 in paper
\cite{Zhao05}, we can see that  BPBC system is readily
accessible experimentally. The transmission spectrum for BPBC is
plotted in Fig. \ref{fig:trans012}. The transmission spectrum is
dominated by the several peaks around the average Fermi energy.
But these overlapping peaks are very broad comparing to those of
BP transmission spectrum and they  form a conductance plateau with
high transmission probability within a broad range of the incident
electron energy. These transmission peaks can be fitted by Lorenz lineshape with 
 the broadening parameters $\sigma$ varying from
0.3 to 0.8 eV, which are significantly larger than that of BP junction 
(typically 0.04 eV). There are two sharp transmission peaks at 1.81 and
2.29 eV above the average Fermi energy on the top of the plateau.
These two sharp peaks are associated with MPSH orbital 48 and 49,
and the four broad peaks correspond to the highest occupied MPSH
orbital (HOMO, orbital 45), the lowest unoccupied MPSH orbital (LUMO,
orbital 46), LUMO+1 (orbital 47) and orbital 50. HOMO and
LUMO+1 come  from the mixture of atomic orbitals from 
 dithiocarbamate groups and BP molecule. HOMO is the bonding state, while
LUMO+1 is the anti-bonding state.

According to the two rules of "robust" transmission spectrum, we
can clearly see that the significant performance improvement is achieved for
BPBC molecular wire junction comparing to  BP junction system.
This improvement is due to the stronger molecule-electrode coupling
which results in the  transmission peak broadening. The shift of peak
position is the  manifestation of the electronic structure of the 
dithiocarbamate group. In fact, in a
model system with the apex Au atom of BP junction substituted by S
atom, we also obtained significantly broadened transmission peaks, 
however the position of these peaks were exactly the same as those 
of BP junction. This result is consistent with the fact that S-Au bond 
is stronger than the N-Au bond, and the $p$ orbital of S  lays too deep
below the electrode Fermi energy  to manifest itself to the transmission spectrum. We also
notice that the conjugation between the S$_2$C group and the end N
atom of BP molecule plays a key role in the conductance
improvement. As indicated in Fig. \ref{fig:MPSH}b, it is the $\pi$
conjugation of the anchoring group which continues the conjugation
from BP molecule directly to the metal.

\begin{figure}
\includegraphics[keepaspectratio,totalheight=16cm]{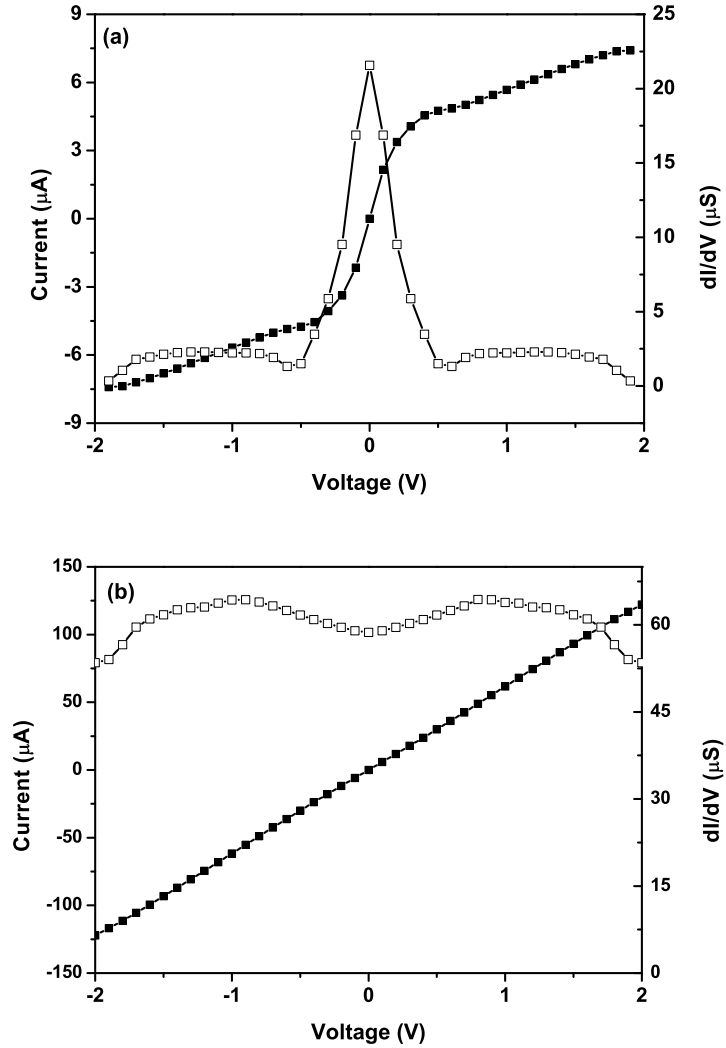}
\caption{Current-voltage (solid squares) and conductance-voltage
(open squares) curves for (a) BP and (b) BPBC molecular wires}
\label{fig:iv}
\end{figure}

Fig. \ref{fig:iv} shows current-voltage (I-V) curve and
conductance-voltage (dI/dV-V) characteristics  of  BP and BPBC
molecular wire junctions. The conductance-voltage characteristics
is computed via numerical differentiation of the I-V curve. The
current dependence on the applied voltage is very different for
BP and BPBC  molecules reflecting the significant difference in
the transmission spectra.  The narrow resonance at the Fermi
energy for  BP molecular wire junctions leads to the steep
increase  of the electric current at the small applied voltage
($\le$0.5 V). The transmission for BP does not have any additional
resonances in the energy range between -1 eV and 1 eV, therefore
once the contribution of the first resonance has been included the
current remains almost constant.  The conductance-voltage
characteristics reflects this behavior of the current-voltage
curve. Conductance reaches its maximum  at near  zero bias voltage
and then it sharply declines to very small values ($<$ 2.5
$\mu$S). The overlap of the peaks in the transmission spectrum for
BPBC molecular wire junction results in the transmission
coefficient, which  is sufficiently large and does not change
significantly with the variation of electron incident energies
from -1 eV to 3 eV (relative to the average electrode Fermi
energy). It leads to the structureless I-V and dI/dV-V
characteristics for  BPBC: current
increases linearly with the voltage bias and the conductance
remains constant within the broad range of bias voltage
variations.

Although DFT plus NEGF represents the state-of-the-art transport
theoretical model, the aim of the direct comparison of experimental and theoretical I-V 
characteristics remains illusive. The conductance enhancement factor (ration between BPBC and BP conductances)
could be less sensitive to particular uncertainties of molecular-electrode interface geometry due the error cancelation. 
 We thus suggest to use the conductance enhancement factor for comparing the results of theoretical calculations 
 and of experimental measurements.
Specifically,  our calculations shows that the conductance enhancement factor at  1.0 V bias, where the conductance of both junctions are very stable (linear regime), is as large as 25. 
We notice that this conductance enhancement factor is consistent with the recent experiment,\cite{Tivanski0598}
in which conductance of biphenyl with thiolate and dithiocarboxylate anchoring group was measured. 
The experimentally measured conductance of biphenyl dithiolate is about 40 times smaller than that of bipyridine.\cite{Tivanski0598,Xu03}
On the other hand, the conductance of the dithiocarboxylate linked molecular wire is enhanced by factor 1.4 with respect to the conductance of
the  thiolate anchored wire.\cite{Tivanski0598} If we assume that the conductance of BPBC and 4,4$^{\prime}$-biphenyl bis(dithiocarboxylate)
are the same within the order of magnitude, then our predicted enhancement factor qualitatively agrees with the existing experimental data. 

\subsection{Asymmetric anchoring}

\begin{figure}
\includegraphics[keepaspectratio,totalheight=16cm]{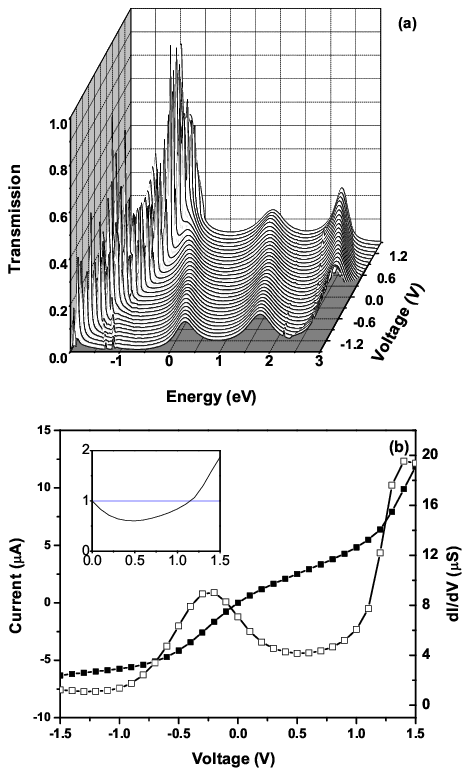}
\caption{
(a) The transmission curves of BPC system at different
bias voltages. (b) The
current-voltage (solid squares) and conductance-voltage (open squares) curves of BPC system.
Inset: rectification-voltage curve (abscissa -- voltage;
ordinate --rectification coefficient). } \label{fig:asymm}
\end{figure}

We use BPC molecular wire junction as a prototype system to
study  rectification of electron transport caused by the
asymmetric anchoring groups, in which the left end of the bipyridine 
molecule is connected
to the gold electrode via N-CS$_2$ linker whereas the right is
attached to the electrode by N-Au bond. It is known experimentally
that the force needed to break the N-Au bond is 0.8 nN, while the
corresponding value for the S-Au bond  is larger than 1.5 nN.
\cite{Xu03a} It indicates that  the bonding between N and the gold
surface is at least twice as weak as Au-S bond. With two S-Au
chemical bonds on the left and with single N-Au bond on the right
such hypothetical molecular wire junction would interact
approximately 4 times stronger with the left electrode than
with the right one.

 The voltage dependent transmission
spectrum is  plotted in Fig. \ref{fig:asymm}a. There are two main
peaks located in the vicinity of the average electrode Fermi
energy. These peaks are broad but their heights are much smaller
than those of  BPBC and BP molecules. The first peak crosses the
average Fermi level energy  and the second peak moves toward the
Fermi energy as the voltage bias increases. The MPSH orbitals for
BPC molecular wire junction are plotted in Fig.
\ref{fig:MPSH}c. Table \ref{tbl:energy} gives the corresponding
eigenenergies. The electron transport through HOMO (orbital
43 in Fig. \ref{fig:MPSH}c) contributes to the first peak. The
other two transmission peaks above the Fermi energy are
associated with  orbitals 44 and 47. Due to the strong coupling
between molecule and left electrode, the transmission
peaks are broad. On the other hand,  because of the asymmetric
electrode-wire coupling, the conjugation of the molecular orbitals
does not continue into the right electrode, which leads to the low 
transmission probability. In fact, the MPSH orbitals are
asymmetric even if we focus on  the region of BP 
molecule. 

The current-voltage characteristic of BPC  is plotted
in Fig. \ref{fig:asymm}b. As is expected from the transmission
spectrum analysis, the I-V curve is asymmetric.  Rectification
coefficient $R(V)=|I(V)/I(-V)|$ is used to quantify the asymmetry
of the molecular wire response on  positive and negative applied
voltage bias. $R=1$ means that there
is no rectification of the electric current. As shown in the inset
of Fig. \ref{fig:asymm}b, for  BPC junction, $R<1 $ when the
$V< 1.15$ V and  $R>1$ for the higher voltages. This means that
the preferential direction for the current swaps with bias
voltage: electrons flow easier from dithiocarbamate linked end
toward nitrogen linked end at low applied voltage bias. When the
voltage becomes larger than 1.15 V the preferential direction
changes to the opposite.

The conductance of BPC molecular wire junction is small at large
negative voltage bias, it grows up as the voltage
increases. It reaches its first maximum ($\sim$ 9 S) at -0.3 V
then decreases to its local minimum ($\sim$ 4  S)
at 0.6 V. The conductance curve shows  five fold increase
within the voltage range of 0.5-1.5 V. This behavior of dI/dV-V
can be explained from the  variation of the transmission spectrum
as a function of the bias voltage. We notice from Fig.
\ref{fig:asymm}a  that the dependence of the transmission peaks
upon the applied voltage bias follows the chemical potential of
the electrode with stronger linkage (left electrode in  our case).
As was pointed out by Taylor \textit{et al.}, \cite{Taylor02} this
is the main reason for asymmetry in I-V characteristics for
symmetric molecules. Electric current is obtained  by the
integration of the transmission spectrum from the left chemical
potential $\mu_L$ to the right chemical potential $\mu_R$
(eq.(\ref{eq:iv})). When the bias voltage varies from 0.0 to -0.25 V, the
transmission peak enter the integration range (from $\mu_L =$-V/2
to $\mu_R$ =V/2). No new transmission peaks contribute to the integral for
the negative voltage bias from -0.25 to -1.5 V. That is the reason
why that maximum appears just below the zero bias voltage in the
dI/dV-V curve. On the other hand, at the positive bias voltage
side, at about 1.5 V, two transmission peaks enter  the range of
the integration and contribute to the integral, which   generates
the inversion of the rectification direction.

\section{Conclusions}
Based upon the combination of DFT and NEGF, we have calculated
the conductance for BP, BPBC and BPC molecular wire junctions. Our
calculations show that dithiocarbamate linking to the Au electrode
leads to the strong molecule-electrode coupling, which is able to continue
the $\pi$ conjugation from the molecule to the metal. The overlap
of the peaks in the transmission spectrum for  BPBC junction
results in the transmission coefficient, which  is sufficiently
large and does not change significantly with the variation of
electron incident energies from -1 eV to 3 eV (relative to the
average electrode Fermi energy).  The electric current increases
linearly with the voltage bias and the conductance remains
constant within the broad range of bias voltage variations for 
BPBC molecular wire junction. Our calculations thus demonstrate
that the conductance of BP molecular wire can be significantly
improved by using dithiocarbamate anchoring group. A conductance
enhancement by factor of 25 in the linear regime is predicted, which is qualitatively
 consistent with the experimental data.
We propose to use
dithiocarbamate anchoring group to make asymmetric molecular wire
junction. We studied the rectification of the electron transport
caused by the asymmetric anchoring groups on the example of 
BPC wire. We show  that  BPC molecular wire junction rectifies
the electric current by factor close to 2. The preferential direction for the current
flow undergoes the inversion at critical voltage 1.15 V.

\section*{ACKNOWLEDGMENTS}
The authors are grateful to M. Gelin, L. Sita, A. Vedernikov and
J. Yang for helpful discussions. Part of the calculations were
performed on computational facilities at the Laboratory of Bond
Selective Chemistry, USTC. This work was partially supported  by
the Mitsubishi Chemical Corporation.

\end{document}